\documentclass[12pt]{article}
\usepackage{graphicx}  
\usepackage{graphics}  
\usepackage{epsfig}
\begin{document}

\title{Photoinduced Phase Transitions}

\author{K.H.Bennemann,\\ Institute for Theoretical Physics, Freie Universit\"{a}t Berlin,\\ Arnimallee 14, 14195 Berlin}
\maketitle

\abstract{Optically induced ultrafast electronic excitations with sufficiently long lifetimes may cause strong effects on phase transitions like structural and nonmetal$\rightarrow$metal ones. Examples are transitions diamond$\rightarrow$ graphite, graphite$\rightarrow$ graphene, non-metal$\rightarrow$ metal, solid$\rightarrow$ liquid and vapor $\rightarrow$ liquid, solid. A spectacular case is photo-induced water condensation. These non-equilibrium transitions are an ultrafast response, on a few hundred fs-time scale, to the fast electronic excitations. The energy of the photons is converted into electronic one via electronic excitations changing the cohesive energy. This changes the chemical potential controlling the phase transition. In view of the advances in laser optics photon induced transitions are expected to become an active area in non-equilibrium physics and phase transition dynamics. Conservation laws like energy or angular momentum conservation control the time during which the transitions occur. Since the photon induced effects result largely from weakening or strenghtening of the bonding between the atoms or molecules transitions like solid/liquid etc. can be shifted in both directions. Photoinduced transitions will be discussed from an unified point of view.}
\newpage
\tableofcontents

\section{Introduction}
Due to advances in laser optics non-thermally affected phase transitions like non-thermal melting etc, see Stampfli et al., resulting from electron excitations due to laser irradiation are of increasing interest \cite{1}. This sheds light on non-equilibrium physics and phase transition dynamics and the role played by conservation laws like energy and angular momentum conservation.
The general physics behind photo-induced phase transitions is illustrated in Fig.1. The potential energy surface (PES)
or thermodynamical potential (Gibbs potential, free-energy) typically separating two phases A and B by a barrier
$\Delta G$ is changed at non-equilibrium resulting from electron transitions changing the populations of the states involved strongly in bonding. If the lifetime of the excited electrons is long enough then the system will respond to the non-equilibrium via phase changes. Important examples are ultrafast structural changes and non-thermal melting, see Stampfli \cite{2}, occurring in covalent crystals, Ge, Si, GeAs, etc.. Due to $sp^3$ population changes and excitations into antibonding states bonding is weakened \cite{3,4}. The opposite case photoinduced bond strenghtening may cause condensation of vapor, non-thermal condensation, or a transition of liquid into solid. Then depending on the energy barrier between the phases the induced phase remains or may change again in accordance with the lifetimes of the excited electrons. Examples of photoinduced bond-strengthening
are the weak van der Waals interactions between Hg-atoms in vapor changing into covalent or metallic ones upon exciting $6s$ electrons into $6p$ states, see theory by Pastor {\it et al.} \cite{5} and experiments by Hensel {\it et al.}\cite{6}, and also the interaction increase between water molecules with weak H-bonding upon electron excitations enhancing the water dipole moments, see W\"{o}ste {\it et al} \cite{7}.

Of course photoinduced effects are also expected from ionization in particular in highly polarizable dielectric material. Note, ionization causes locally quasi strong ionic bonding.

As physically expected the photoinduced response
must be shorter in time than the lifetimes of the excited electrons. This was noted and discussed already by Hensel {\it et al.} when explaining the different response of Hg and Cs vapor to light \cite{6}.

In general, the chemical potential, the thermodynamical potential per particle  $\mu(t)$  controls the phase stability and phase transition and is in
general changed if bonding varies due to electronic excitation \cite{8}. This is the origin of non--thermal photoinduced effects on phase transitions. Of particular interest are, for example, the effects on supercooling or superheating or supersaturation. If no energy barriers stabilize the induced phase, the nonequilibrium state will relax in accordance with the lifetimes of the excited electrons. Of course, as noted already the time needed for the induced transition must be faster than the lifetimes of the electronic excitations.

In Fig.1 the physics of photoinduced phase transitions is illustrated. Note, after relaxation of the excited electrons the system may return or also not to the original state. It is generally important to achieve optically for example not only the transition crystalline $\rightarrow$ amorphous,
but also amorphous $\rightarrow$ crystalline etc..

In Fig.2 we illustrate photoinduced effects on non--thermal melting in covalent crystals. Note, the transition is ultrafast
and occurs during 100 fs or so \cite{2}.
\begin{figure}
\centerline{\includegraphics[width=.6\textwidth]{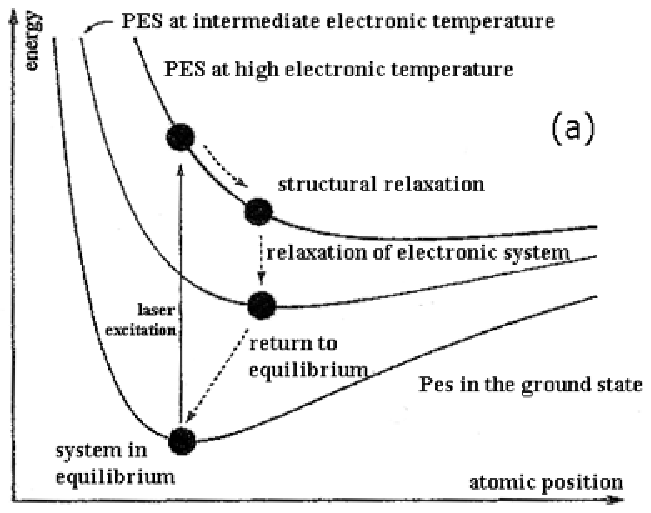}}
\vspace{1cm}
\centerline{\includegraphics[width=.5\textwidth]{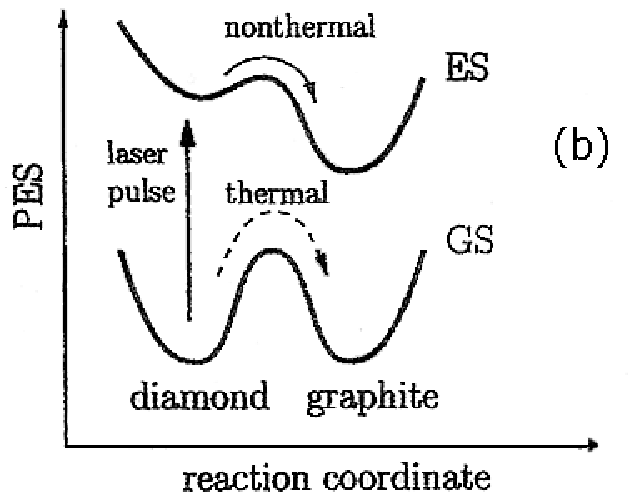}}
\caption{ (a) Illustration of the physics behind nonthermal and photoinduced phase transitions. For a detailed discussion see Garcia, Jeschke \cite{3,4}. Note, irreversible transitions may occur. After relaxation of excited electrons system may not return to original state. (b) Illustration of photoinduced changes of the potential energy surface (PES). Electrons are excited out of the ground state (GS) by a laser, for example, into the non--equilibrium state ( ES ) with a smaller energy barrier between two phases A and B thus favouring the transition A$\rightarrow$B. This may refer to ultrafast nonthermal melting of covalent crystals, to the transition diamond$\rightarrow$graphite or photoinduced condensation of supersaturated or supercooled vapor.}
\end{figure}

In Fig.3 we illustrate for example what is expected for supersaturated or undercooled vapor. Note, in particular close to an equilibrium phase
boundary the electron excitations may induce for saturated vapor a transition vapor$\rightarrow$liquid, but also directly vapor$\rightarrow$solid. In case of photoinduced
enhancement of dipole coupling the latter transition results from the energetically favored ordering of the dipole moments in the solid as compared to the liquid, see water vs. ice .
\begin{figure}
\centerline{\includegraphics[width=.6\textwidth]{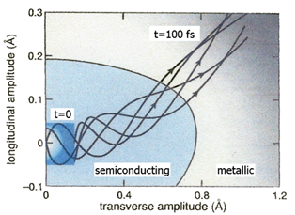}}
\caption{ Results by Stampfli {\it et al.} \cite{2} for the ultrafast phase transition semiconductor$\rightarrow$metal and nonthermal melting. This phase transition results from exciting valence electrons into the conduction band thus causing a weakening of the bonding and a vanishing gap between conduction and valence band (metallic state). Various trajectories as a function of the induced transverse acoustic (TA )-- and longitudinal optical (LO )--atomic distortions are shown. Note, the TA-- and LO--atomic displacements dominate the induced structural changes. The results refer to Si, but are similar for other covalent crystals. Note, Si melts after a time of about 100 fs. The photoinduced transition depends of course on the density (fraction of excited electrons) $\xi$ of electron--hole excitations. Results refer to $\xi\approx$ 0.15.}
\end{figure}

The thermodynamical potential $\Delta G$ determines the phase transition, see Fig. 1 and Landau, Kubo {\it et al.} \cite{8}. In case of a transition A$\rightarrow$B via forming nuclei of the new phase B
the phase transition rate $\Gamma$ for forming nuclei is controlled by $\Delta G$ and it is
\begin{equation}
\Gamma_{A\rightarrow B} \propto  \exp -(a\Delta G) ,
\end{equation}
with $\alpha= 1/kT$ if the transition is a thermal one. Hence, there is a sensitive dependence on changes in the potential barrier $\Delta G$ due to electron excitations.
\begin{figure}
\centerline{\includegraphics[width=.45\textwidth]{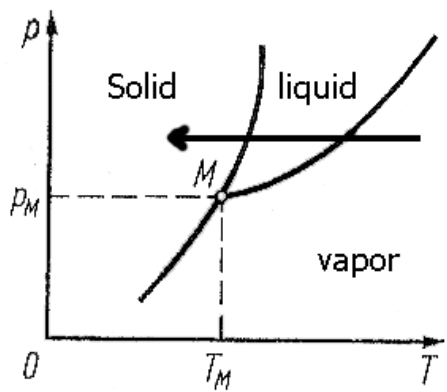}}
\caption{ Pressure vs. temperature Phase diagram for transitions between vapor, liquid and solid. A state is thermodynamically stable if the chemical potential $\mu(t)$ is minimal. Electron excitations change bonding and thus $\mu(t)$. Moreover supersaturation or supercooling is affected by the electron excitations, since these may favor the formation of nucleation centre necessary for condensation. The photoinduced condensation of supercooled vapor close to the triplepoint M is illustrated. The arrow demonstrates supercooling. Note, the photoinduced condensation of the supercooled vapor occurs at the temperature $T$ (top of arrow). Note, close to M, if $(T-T_0)$ is large enough, one gets a transition vapor$\rightarrow$solid. Both supercooling $(T_0-T)$ and supersaturation $(p-p_0)$ is getting smaller for photoinduced bond strenghtening. An indication of the latter is the latent heat.}
\end{figure}

Note, for transitions via precursor fluctuations involving small nucleation centres with N particles of the growing new phase surface energy $\sigma $ must be included in the thermodynamical potential $G$. This is for example the case for small liquid droplets in vapor, for small crystallites in melt, etc.

Generally, the difference of the thermodynamical potential $\Delta G$ between phase A and B is for const.volume given by \cite{6,7,8}
\begin{equation}
                 \Delta G = -\Delta S\Delta T -N\Delta\mu +\sigma \Delta s +\ldots .
\end{equation}
Here, $\Delta S$ is the entropy difference and $\Delta \mu$ the one of the chemical potential and $\sigma$ is the surface energy (per unit area) of the interface of A and B. The first term may be rewritten as $(N q /T)\Delta T$, where $q$ is the molecular latent heat. Note, the latent heat should take into account the change in binding due to the electron excitations.
$\Delta T= T_0-T$ gives the range of supercooling, temperature $T_0$ refers to the temperature at the phase boundary (or to the starting temperature) and condensation occurs at T when the new phase nucleus arising from fluctuations has reached its critical size and continues to grow.
The chemical potential change $\Delta\mu$ involves strenghtening or weakening of the bonds between the particles. It is \cite{8}
\begin{equation}
                 \Delta\mu = T ln (p/p_0) + \Delta \mu{'},
\end{equation}
where the last term describes the change of the bonding of the particles. $p_0$ is the pressure at the thermodynamical phase boundary. Summarizing,
Eq.2 can be rewritten as
\begin{equation}
               \Delta G = -(Nq/T)\Delta T - NT\ln(p/P_0) - N\Delta \mu{'} + \sigma \Delta s +....
\end{equation}

Clearly, $\Delta G= \Delta G_0 + \Delta G_{ph.}$ and the photoinduced contribution $\Delta G_{ph.}$ depends on the light fluence F, since the probability to absorb a photon and to excite an electron should be proportional to F. Hence, approximately one has ($\ln\Gamma \propto F/kT$)
\begin{equation}
               \Delta G_{ph.} \propto F.
\end{equation}
From this one gets approximately, if $\Delta G_{ph.}$ is not too large, for the rate of forming nuclei
\begin{equation}
               \Gamma \propto F/kT + ... .
\end{equation}
The enhancement of the rate $\Gamma$ due to photon absorption is given by
\begin{equation}
               \Gamma = \Gamma_0 \exp (-\Delta G_{ph.}/kT).
\end{equation}

Note, in case of phase transitions involving
small nucleation particle clusters, one assumes for the nucleation centre $N \sim$ volume  and for calculating the surface energy $\sigma$ (per unit surface area) small spherical particles. For the spherical nucleation centres it is $\Delta s=4\pi r^2$. The particle density is $\rho = N/V$.

In case of phase transitions via (spherical) nucleation clusters having radius r one gets approximately
\begin{eqnarray}
 \Delta G \simeq &-&\!\!\!(4\pi/3)r^3(q/T)\rho \Delta T -(4\pi/3) r^3\rho T\ln(p/p_0) - (4/3)\pi r^3 \rho\Delta \mu{'}\nonumber \\ &+&\!\!\! 4\pi r^2\sigma
                              + n(e^2/2\epsilon r) (1-1/\epsilon) + \ldots.
\end{eqnarray}.
Here, q is the molecular latent heat, $p_0$ refers to the equilibrium pressure at a flat surface and $(p-p_0)$ gives the range of saturation. $S=p/p_0$ is called supersaturation, $\rho$ is the particle density and $\sigma$ is the energy per unit surface area of the liquid droplet or crystallite in the melt. $\Delta \mu{'}$ is the change of binding energy per particle. The photoinduced effect is proportional to the number of excited atoms or molecules in the nucleation cluster, its fraction is denoted by $\xi$, and may be determined by the absorption cross section and the light intensity, the fluence F.

Note, in the third term in above equation we assumed for simplicity an uniform change in the nucleus, that all particles in the nucleus have been excited. If the fluence is not too large this will not be the case and the term should be calculated more correctly. For example, one may assume that the particle at the centre of the nucleus absorbed a photon and caused electron excitation changing the cohesion (The situation corresponds to the one of an impurity in a liquid or solid, or in case of several excited particles in the nucleus to changes in cohesion like in an alloy).

The last term refers to effects due to ionization which may be caused also by photon absorption. Thus, n ions with charge e may be present in the nucleation centre. Again, n depends on F. Screening of the charges is taken into account by $\epsilon$ which is the dielectric function of the material \cite{8}. Again we assume for simplicity that ions are nearly located close to the centre of the nucleus.

It follows from the above equation for $\Delta G$ that photoinduced increased cohesion decreases the critical size $r_k$ beyond which the nuclei grow and the new phase gets stable. Denoting by $V_k$ the volume of a critical nucleus one expects approximately from above theory
\begin{equation}
            V_k = (V_k)_0 - \alpha F,
\end{equation}
where the photoinduced change of the critical fluctuation size $(V_k)_0$ should be observable using a fast measurement (Mie scattering).

Furthermore, one gets for critical supercooling (when nuclei have reached the critical radius $r_k$) approximately
\begin{equation}
             \Delta T = (T/q) ( 2\sigma/ r_k + \Delta \mu{'} +...),
\end{equation}
hence $\Delta T$ decreases for increasing excitations, fluence F and photoinduced increased cohesion.

For critical supersaturation one gets
\begin{equation}
             \ln S = 1/T ( 2\sigma/r_k + \Delta \mu{'} +...).
\end{equation}
Note, $\ln S$ decreases with temperature T and becomes smaller for increasing cohesion, F.

For increased binding it is $\Delta \mu{'} < 0$ and hence as expected critical supercooling $(T_0 - T)$ and supersaturation $(p - p_0)$ decreases upon light absorption. Of course, the latent heat q should reflect also increased cohesion ($\Delta \mu{'} \propto F$).

Corresponding  effects occur for decreased cohesion, $\Delta \mu{'} > 0$, resulting for excitations into antibonding states.

In above Eqs. it is straightforward
to write down explicitly the effects due to ions or enhancement of dipolar coupling.

A particular interesting case is the photoinduced increase of the interaction between molecules due to dipole moments. Electron excitations can change, enhance the dipole moment. Thus the cohesion can increase and affect vapor condensation etc.. This applies to photoinduced water condensation. Dipole interactions must be taken into account in the term $\Delta \mu{'}$.

Induced dipole moments contribute to $\Delta \mu{'}$ a term increasing with p, where p denotes the dipole moment. In polarizable systems it is $p\sim E$ where the dipole moment is induced by an electric field $E$.

For calculating $\Delta \mu{'}$ due to dipole coupling one must take into account the ordering of the dipole moments, see for example water vs. ice \cite{9,10}.
The interaction between molecules at distance r with dipole moments p is given by the potential energy $U=-r^{-6}[(2/3)p^{4}/kT + ap^2 +...]$,
if the first term is much smaller than kT, see Keesom. Note, if at low temperature T dipole coupling is much larger than kT one must
replace the first term in U by $2p^2/r^3$ \cite{9}. It is
\begin{equation}
           \Delta G_{dipole} \simeq \int^r 4\pi \rho^2 d\rho U_{dipole}.
\end{equation}
Then assuming at the centre of the nucleus a photoenhanced dipole moment one gets
\begin{equation}
           \Delta G_{dipole} \sim (8\pi p^2 p_0^2/kT)1/r^3 , \mbox{at high T},
                             \sim (8\pi p^2) 1/r,   \mbox{at low T},
\end{equation}
in contrast to proportional to $p^4$ etc. for a uniformly excited nucleus.
Note, in water the dipole coupling may contribute to more than seventy percent of the cohesion.

The dipole coupling yields at higher temperatures to the thermodynamical potential $\Delta G$ a contribution $a(p)p^2(p_0)^2 +...$, where the coefficient a increases with pressure in accordance with the intermolecular distance dependence of the coupling. At low temperature one gets a contribution to $\Delta G$ proportional to  $a{'}(p)p^2$. The dipole moment $p_0$ refers to not excited particles.
One may write the dipole moment as $p= p_0 + \Delta p$, where $\Delta p$ denotes the photon induced change of the dipole moment. Approximately, $\Delta p \propto F$.

Thus, similarly as in the case of radiation producing ionization photoinduced enhancement of the dipoles and their coupling may cause for example strong condensation of water (or even ice in case of supercooling or supersaturation). It is using previous discussion
\begin{equation}
                  \ln \Gamma \propto - (\Delta G)_{dipole},  (\Delta G)_{dipole} \propto p^2 (p_0)^2 / kT +... \mbox{at higher T}.
\end{equation}
It is interesting that at low temperatures T the behavior changes and $\ln \Gamma \propto p^2 +...$.

Generally one gets that the rate of condensation $\Gamma$ increases for increasing fluence F, likely first linearly. The condensation rate $\Gamma$  increases also with supersaturation S or supercooling. It would be interesting to observe effects reflecting the number of particles excited within the nuclei, fraction of excited particles $\xi$.

In Fig.4 we sketch the resulting $\Delta G$ for nucleation, see previous discussion by Hensel {\it et al.} \cite{6}.
\begin{figure}
\centerline{\includegraphics[width=.5\textwidth]{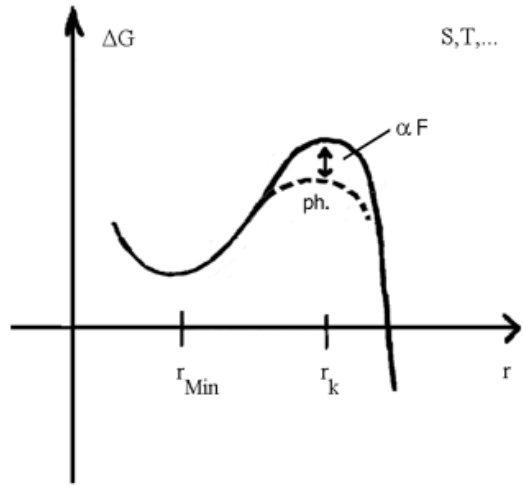}}
\caption{Illustration of the general behavior of the thermodynamical
potential G for a phase transition resulting from fluctuations involving nucleation, for example droplet formation. $\Delta G$ depends on $S=\frac{p}{p_0}$ and temperature T. Here, r denotes the radius of the nucleus, for simplicity assumed to be spherical. Note, the phase fluctuations accompanying the phase transition may consist of droplets which act as precursor of the new phase. At the critical radius $r_k$ vapor condensation or crystallization occurs. The rate of nucleation $\Gamma$ is determined by the potential barrier $\Delta G_k$ which results from the interplay of surface energy and changes of cohesion. Note, $\Gamma$ could be observed by Mie scattering. Photoinduced electron excitations may lower this barrier $\Delta G$. The dashed curve refers to photoinduced reduction of the potential barrier and which is approximately proportional to light fluence F.}
\end{figure}

Note, the photoinduced increase of the dipole moments is expected to be an important source for photoinduced water condensation of vapor. The photoinduced transition vapor$\rightarrow$ice may be preferred for supersaturated or supercooled vapor, for energetic reasons, see Zhadanov, Pauling, {\it et al.} \cite{9,10}. The ordering of the dipole moments in ice yields stronger binding. Thus, supercooled or supersaturated water vapor may exhibit photoinduced condensation to ice.

In the following we apply the discussion to photoinduced
\begin{enumerate}
  \item Structural Transitions,
  \item Non-thermal melting of covalent crystals,
  \item Condensation of vapor
  \item Condensation of water.
\end{enumerate}

\section{Applications}

The photoinduced electronic excitations change the bonding and consequently the chemical potential and this affects the phase transitions. Generally one expects that the number of
electron excitations (or of excited atoms or molecules) is proportional to photon absorption, to the laser fluence F, at least approximately. As a consequence one expects for the energy barrier $\Delta G_k = (\Delta G_k)_0 + (\Delta G_{k}){'}_{ph.}$ and $(\Delta G_k){'}_{ph.}  \sim F +...$, see Fig.4.

\subsection{Photoinduced Structural Changes}

Interesting responses to laser irradiation are expected for systems where the electron excitations change the bond character like in covalent solids with $sp^3$ bonds changing to $sp^2$ ones or weaken the covalent bonds strongly enough that melting occurs. In general photoinduced excitations
from bonding into antibonding states may cause structural transitions like graphitisation of diamond, nonthermal ultrafast melting and ablation at surfaces. Ablation of an intact graphite plane may result at the surface of graphite due to laser irradiation. Repopulation of s and $p_x$, $p_y$ and $p_z$ states causes such responses. Thus, graphene can be produced optically. This is a remarkable example of photoinduced transition.

One may use the Boltzmann equation to calculate the population of the electronic states by the excited electrons. The cohesive energy at non--equilibrium is given by, see Stampfli {\it et al.} \cite{2},
\begin{equation}
     E = \int_{-{\infty}}^{\varepsilon_F} d\varepsilon \varepsilon N(\varepsilon, T_{el}) f(\varepsilon, T_{el}) + E_r({r_{ij}}) +...,
\end{equation}
where N is the electronic density of states, f the Fermi distribution function and $E_r$ the repulsive energy between atoms at distances $r_{ij}$. $T_{el}$ is the temperature of the hot electrons. Then the dynamics of the atomic displacements at nonequilibrium follows approximately from the (classical) equation
\begin{equation}
     M {\ddot{u}_i(t)} = - \frac{\partial}{\partial u_i} E_b (u_i, f, T_{el}).
\end{equation}

As discussed by Garcia {\it et al.} this analysis can be improved using a Lagrangian formalism which also applies if photoinduced transitions are accompanied by volume changes \cite{3,4}. Thus potential energy surfaces (PES) and nonequilibrium phase transitions are calculated.

In particular this analysis can be used to determine photoinduced structural changes in semiconductors. In covalent bonded semiconductors the photoinduced nonequilibrium state resulting from electron excitations is then characterized by the fraction $\xi$ of electrons excited across the energy gap between valence and conduction band. Excitations into antibonding states weakens cohesion. Clearly, transitions $sp^3\rightarrow sp^2$ destroy covalent bonding. The characteristic band gap in the semiconductors decreases. Closing of the gap between conduction band and valence band yields metallic behavior.

In view of this, clearly crystals with strong excitonic behavior are expected to exhibit strong photoinduced effects. Photoinduced creation of electron--hole pairs in graphene might be particularly interesting. To study this one may use the same analysis as for the graphitization transition and for ablation \cite{3,4}.

Photoinduced changes of the bonding are expected for rare--earth and generally for crystals (non--metals) with weak bonding due to a band gap as is typically the case for a filled outer electron shell. Vapor of mercury etc. might be an example. Then electron excitations across the band gap could change weak van der Waals like bonding to stronger metallic one.

\subsection{Photoinduced Strenghtening of Bonding}

Of course, due to electron excitations interatomic or intermolecular interactions, binding may also increase. This may be the case for electron excitations out of a filled electronic shell, see for example Hg, Ba, Ce with $6s^2$, Zn with $4s^2$, lanthanides...etc.. If these excitations live long enough the nonequilibrium population of the electron states may change the phase. In particular photoinduced increased cohesion of the nonequilibrium state should increase the melting temperature and favor condensation of vapor.

Regarding phase diagrams, note the Clapeyron--Clausius equation \cite{8} gives $\frac{dp}{dT}= \frac{q}{T(v_2 - v_1)}$, where $v_2$ and $v_1$ refer to the volume in the two phases 1 and 2. Here, phase 2 may be the gas one and phase 1 the liquid one. Photoinduced effects result from changing the latent heat q (for example of vaporization or melting,..) due to electronic excitations. Increased bonding increases q and consequently for example the temperature at which melting or evaporation occurs. As known \cite{8} one gets for the transition liquid$\rightarrow$vapor
\begin{equation}
      p_s = \exp(-q/kT)
\end{equation}
for the phase boundary. Here, one may write $q= q_0 + \Delta q$ and approximately $\Delta q \propto F$. Increased bonding increases also the temperature for evaporation (or melting). This should occur within very short time, for example within 100 fs or so. Note, electron excitations, hot electrons raise also the electron temperature and this must be taken into account.

Photoinduced transition of a metastable state into a stable one deserves special attention. In particular supersaturation and supercooling etc. should be affected by photoinduced electron excitations. One expects photoinduced condensation of water and of mercury etc., since the excited electrons increase cohesion, at least for some time depending on the lifetimes of the excited electrons. In vapor of Hg the weak van der Waals interactions change to covalent or even metallic interactions due to $6s^2\rightarrow 6s^1p^1$ transitions \cite{5,6}. In water dipole coupling may increase strongly upon electron excitations and thus enhance the weak intermolecular coupling of the vapor.

Then, assuming that vapor condensation or crystallization results from fluctuations consisting of nucleation centres (liquid droplets or crystallites) one may rewrite Eq.(4) as
\begin{equation}
     \Delta G \simeq -(4/3)\pi r^3 \rho (q/T )\Delta T - 4\pi r^3 \rho T \ln S - (4\pi/3)r^3 \rho \Delta\mu{'} + 4\pi r^2 \sigma +...,
\end{equation}
where $S=p/p_0$ is the supersaturation (and where $p_0$ refers to the equilibrium pressure of a flat surface) and where the degree of
supercooling is given by $\Delta T= T_0 -T$. Here, $T_0$ is the temperature at which the two phases coexist thermodynamically, at equilibrium. Note, the (molecular) latent heat should refer to the case where electron excitations are present. One may write again $q= q_0 + \Delta q$, where $\Delta q$ is due to the electron excitations and is approximately proportional to the fluence F. Photoinduced condensation (or crystallization) results as discussed already from a decrease in the barrier of $\Delta G$ \cite{6}.

One expects physically that photoinduced increase of cohesion reduces the critical size, radius $r_k$ of the nucleus (droplet in vapor or crystallite in melt). It is approximately (from $\Delta G{'}= 0$)
\begin{equation}
             r_k = \frac{2\sigma}{(q/T) \Delta T + \rho T \ln S - \Delta \mu{'}},
\end{equation}
hence $r_k$ decreases if $\Delta \mu{'}$ is negative.

Furthermore, for example for constant pressure and from $\Delta G{'} = 0$  one gets for the temperature T at
which condensation of supercooled vapor occurs
\begin{equation}
     T_0 - T \simeq (2\sigma T_0)/q \frac{1}{r_k} + \frac{T}{q}\Delta \mu{'}.
\end{equation}
Here, $r_k$ is the critical size of the nucleation centres (obtained from $\dot{\Delta G{'}}= 0$). Note, for $r > r_k$ condensation of vapor (or crystallization of melt) occurs, since the nucleation centre grow. As noted already one may assume approximately  $\Delta q \propto F$, where F refers to the fluence of the laser light. Obviously, if $\Delta \mu{'}$ is large enough $\Delta T$ gets smaller.

Similarly one gets for supersaturation \cite{8}
\begin{equation}
      p - p_0 \approx 2\sigma / r_k + \Delta \mu{'} .
\end{equation}
Here, the critical size of the (liquid or crystal) nucleus $r_k$ decreases for increasing cohesion and thus the supersaturation $\Delta p$ decreases if $\Delta \mu{'}$ is large enough (note,
$p_0$ refers to a planar interface of the two phases).

In summary, both critical supercooling and supersaturation decrease with increasing light fluence F.

In Hg nanostructures and thin films the gap between the 6s states and 6p states may be small enough so that a transition nonmetal$\rightarrow$metal may result due to electron excitations into the 6p states, see Pastor {\it et al.} \cite{5}. Then photoinduced effects get stronger. The increase in the coupling between the Hg atoms in the condensed nuclei can be related to the behavior of Hg clusters. Note, clusters and nuclei having about 15-20 atoms should have a nearly closed s--p "band gap", optical measurements should reflect this, also latent heat.

The dynamics of the phase transition is controlled by energy--, angular momentum--conservation (see magnetic effects) etc. Then, if in supercooled
melt there is a crystal nucleus with $\mu_s < \mu_l$ it grows for some time, but the whole melt will not crystallize if heat cannot be removed (transferred) fast enough. For crystallization of the whole melt heat must be removed and this may take some time (ps or so) \cite{9}.

Photoinduced condensation of water vapor may result from increasing the relatively weak interaction amongst the water molecules via laser irradiation
and due to intraatomic electronic excitations enhancing the dipole moment of the water molecules. Then one must add explicitly to $\Delta G$ the term
due to dipole--dipole coupling ($\sim \frac{p_1^2p_2^2}{r^6}+...$, $p_1$ may be the photo enhanced dipole moment and $p_2$ the surrounding dipole moments, which are changed due to $p_1$). After averaging over the angular dependence of the dipole coupling one gets approximately (Keesom)
\begin{equation}
      \Delta G_{dipole} \simeq -a(p)(4\pi/3)r^3 [ p_1^2 p_2^2/kT +...], \mbox{large T},
\end{equation}
where p is the average dipole moment of the water molecules, $\varrho$ the density of water and the parameter a depends on the intermolecular distance. Note, approximately $p_1 \propto F$. At low temperatures one gets after averaging over the angular dependence of the dipole coupling
\begin{equation}
      \Delta G_{dipole} \simeq -a{'}(p)(4\pi/3)r^3 [p_1p_2 +...], \mbox{low T}.
\end{equation}
The structure is different for ice and water, see Pauling and new studies \cite{9,10}. In the above Eqs. contributions due to induced dipole moments
via strong electric fields are only indicated by dots.

Note, the increase of the dipole moment p results from changing upon photon absorption the hybridization and population of the 1s and 2p states in $H_2O$. This might lead to a considerable strenghtening of the bonding between the water molecules. Depending on the size, radius r of the condensed nucleus and light fluence F a critical nucleus might involve one or more photoexcited molecules. Correspondingly the cohesion and latent heat q increases.

Note, in highly polarizable material a strong electric field E causing $p= \alpha E$ is expected to increase also the dipole coupling. Then, $\Delta p \propto E$.

One may write generally $p= p_0 + \Delta p$, where $\Delta p$ is photoinduced. Approximately, one may assume again $\Delta p \propto F \propto \xi$. Here, as before $\xi$ is the fraction of excited atoms or molecules.

Note, this photoinduced strenghtening of the dipolar coupling may cause a strong condensation of water in saturated vapor and also for supersaturated water vapor at sufficiently low temperatures ($T\leq q^{'}$, where $q^{'}$ refers to the latent heat for water$\rightarrow$ice) one might get a transition to ice. Here, $q^{'}$ should take into account the increase of the dipole moment.

Of course, ionization may play also a role and strengthen the intermolecular coupling ($\Delta G \propto \frac{e^2}{2\epsilon r} + bep/r^3 +...$).

Since the probability that a phase transition precursor nucleation centre is formed is close to the thermodynamical phase boundary given by, see Eq.1, ($w\sim \Gamma$)
\begin{equation}
      w \simeq A \exp-{\frac{\Delta G}{kT}},
\end{equation}
already relatively small increases in the dipolar coupling between the water molecules may cause strong condensation. The rate of forming condensation clusters is related to the laser fluence F. Note, $\ln w \sim \ln \Gamma \propto F$ or approximately w and $\Gamma$ are proportional to F.

Of course, the electron excitations are controlled by light frequency and the photoinduced effects depend on the lifetimes of the excitations. In principle, the water clusters can be observed by Mie scattering (profile, in particular backward one)\cite{7}.

The enhancement of the probability or rate of condensation due to photon excitations is in accordance with the previous equations given by
\begin{equation}
     \Gamma = \Gamma_0 \exp (\frac{\Delta G - \Delta G_0 }{kT}),
\end{equation}
where $\Delta G_0$ refers to the case when no photoexcitations are present. Thus, one gets approximately for supercooling the enhancement
\begin{equation}
     \Gamma \propto \exp [(\frac{\Delta q}{T^2})\Delta T + ...] .
\end{equation}
\begin{figure}
\centerline{\includegraphics[width=.65\textwidth]{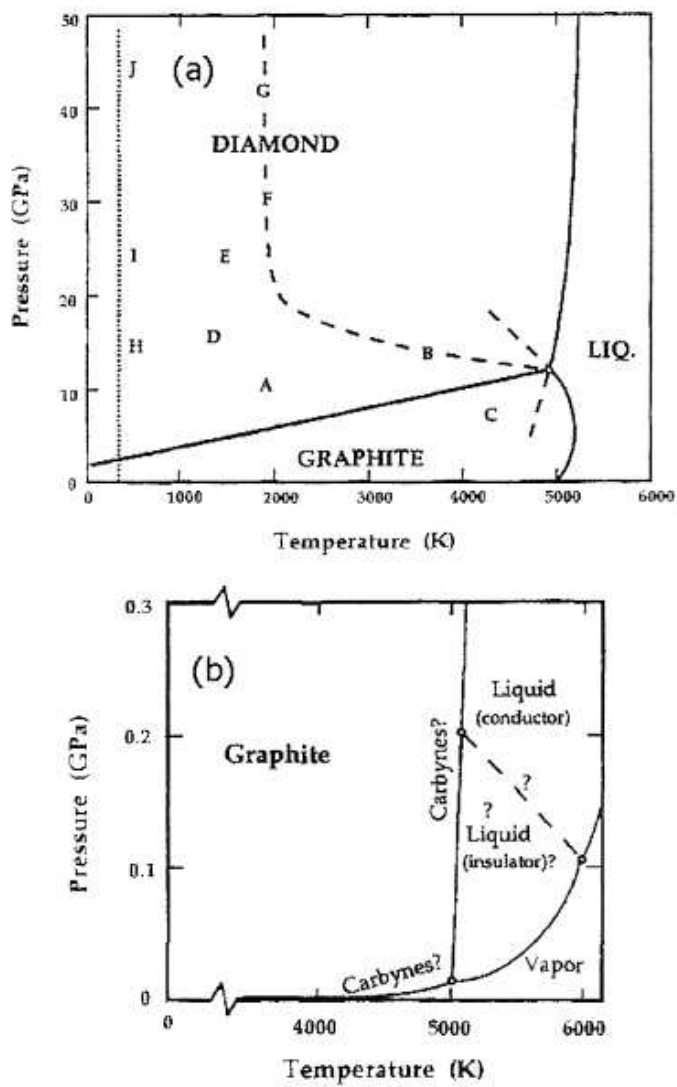}}
\caption{(a) Pressure vs. temperature phase diagram observed for carbon. The solid lines refer to equilibrium, B and C refer to very fast transition of graphite to diamond and diamond to graphite, respectively. For a discussion see Jeschke \cite{4}. These results referring to equilibrium
thermodynamics should be compared with nonequilibrium behavior. (b) Phase diagram for lower pressures (? indicates controversial issues). Carbynes may refer to phase consisting of linear molecules. Supercooling and superheating may exhibit particular interesting behavior. Of great interest is also to explore how to get the transition graphite$\rightarrow$diamond at nonequilibrium, possibly via inducing small diamond crystals which grow rapidly before relaxing into graphite.}
\end{figure}

It is of interest to give in more detail using standard thermodynamics \cite{8} the expression for the formation probability of a nucleus in (slightly) superheated or supercooled phase. It is \cite{11}
\begin{equation}
          w\sim \exp{-\frac{16\pi\sigma^3 T_0^2}{3T\rho^2 [q (\Delta T) + T_0\Delta \mu']^2}} +... ,
\end{equation}
where T may be written as $T_0 - \Delta T$. Using then Clapeyron--Clausius $\Delta p = \frac{q}{T_0(v_1 - v_2)} \Delta T$ one may express $w$ also in terms of $\Delta p$. The molecular volumina  $v_1$ and $v_2$ refer to the metastable phase and the nucleus, respectively \cite{8,9}.

Note, the phase transition via precursor fluctuations consisting of nuclei refers to sufficiently pure substances. The nucleation due to impurities etc. is neglected here for simplicity. Also if for example saturated vapor is in contact with its liquid and the interface is planar, then condensation of the vapor occurs without nucleation and one has no supercooling. (Similarly for such interface no superheating occurs). Formation of liquid nuclei within a crystal if internally heated is possible if surface is kept below melting temperature \cite{8}.

Of course, one gets also photoinduced changes of phase transitions without nucleation, see previous discussion. If cohesion increases due to electron excitations caused by photon absorption, then one expects that ultrafast melting occurs for at higher temperatures

\section{Results}
   Results for the response to laser induced electron excitations due to weakening or strenghtening of interatomic or intermolecular bonding are pre-
   sented. Note, of particular interest are phase transitions like vapor/liquid, crystalline/amorph, etc. There are many problems, for example
   optically induced transitions diamond/graphite, graphite/graphene and in general problems regarding agglomeration physics. Such optical studies
   may help to shed light on non-equilibrium physics and time resolved ultrafast phase transitions. For example, time dependence of crystal growth,
   at equilibrium vs. non--equilibrium.

\subsection{Bond Weakening}

   First we present results typical for photoinduced weakening of chemical bonding. Examples are structural changes in covalently bonded semiconductors and ablation at surfaces.

   In Fig.2 typical results for photoinduced transitions of covalently bonded semiconductors are shown. Depending on the light fluence F or concentration $\xi$ of electron excitations, excitations into antibonding states, the covalent structure changes very fast, within about 100 fs,
   and conductivity increases, since the gap between conduction and valence band closes.

   \begin{figure}
   \centerline{\includegraphics[width=.8\textwidth]{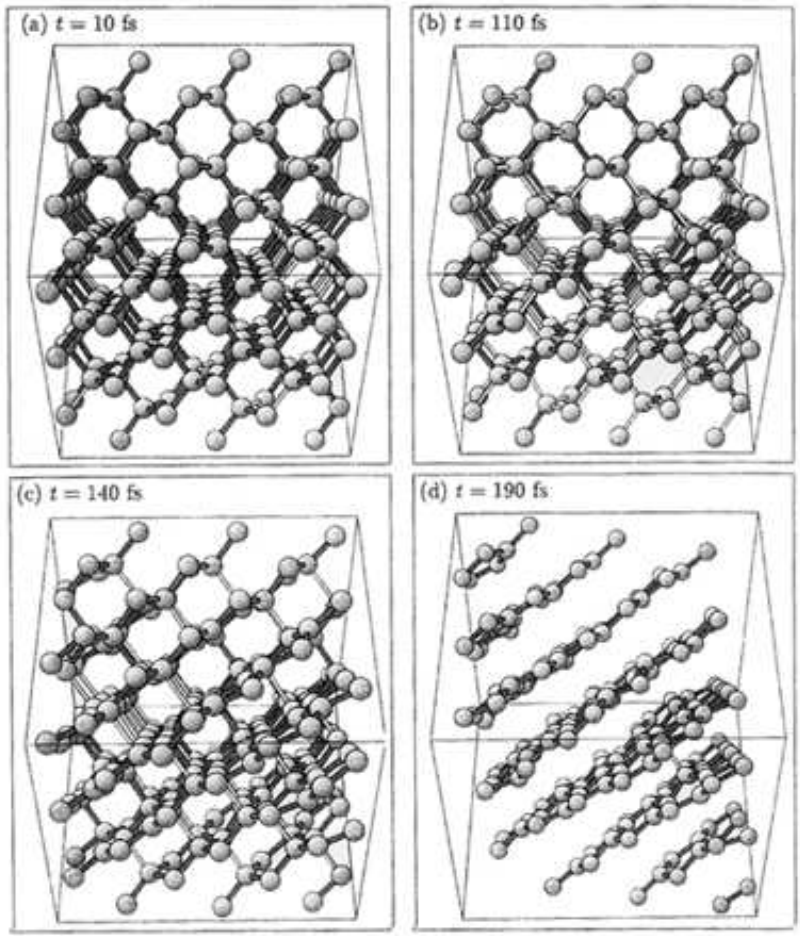}}
   \caption{Ultrafast nonthermal graphitization of a (100) diamond crystal. Note, 1.1 eV/atom is absorbed from the laser pulse of 20 fs duration. The transition is due to electron excitations. For details of the calculations see Jeschke \cite{4}. Note, while for diamond $p_z$ orbitals are important for structure and cohesion, for graphite $2p_z$ states are not essential for binding. As a consequence the diamond minimum in PES gets weaker and the one for graphite does not change much.}
   \end{figure}
   In Fig.6 the photoinduced very fast transition diamond$\rightarrow$graphite is shown, results are obtained by Garcia and Jeschke \cite{3,4}.
   The graphitizataion occurs, since the photoinduced repopulation of the s, p$_x$, p$_y$, p$_z$ states lowers essentially the minimum of G or PES of diamond, see Fig.1(b). The minimum of the PES for graphite remains largely unchanged.

   In Fig.7 important results by Jeschke, Garcia \cite{3,4} are given for photoinduced ablation at the graphite surface.
   Note, the coherent ablation of a graphite plane, quasi optical generation of graphene.
   \begin{figure}
   \centerline{\includegraphics[width=1.\textwidth]{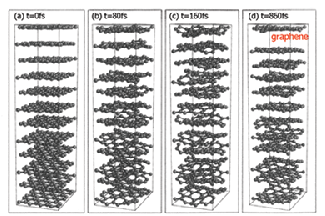}}
   \caption{ Typical results for nonthermal ablation at graphite surface. Note, 2.4 eV/atom is absorbed from the laser pulse acting for 20 fs. The light density is below the critical one destroying graphite planes. After about 160 fs the light energy is transferred to the atoms causing strong coherent vibrations of the surface planes and ultimately even a detachment of whole intact graphite planes. This optical production of graphene first calculated by Jeschke {\it et al.} \cite{4} was later observed.}
   \end{figure}

   In Fig.8 the mechanism for the coherent ablation of a whole plane is illustrated. This may have more general validity for ablation processes. In some cases depending on the crystal structure coherent ablation may depend on the laser pulse shape, duration of the pulse and wavelength and light  polarization.

   Of interest is also to study the optical response of graphene or a few layers of graphene. In particular how is the momentum dynamics, the transfer of the photoinduced momentum, how long does it take to distribute transverse one into longitudinal one. etc.. This sheds light on the stability of graphene, its wavy like structure etc..
   \begin{figure}
   \centerline{\includegraphics[width=.7\textwidth]{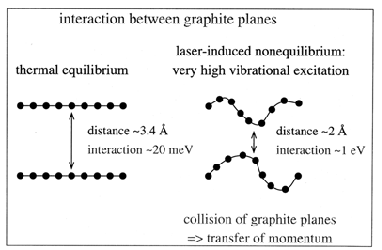}}
   \caption{Illustration of the mechanism responsible for the photoinduced formation of graphene, see calculations by Jeschke {\it et al.} \cite{4}.
   The photoinduced vibrational motion of the graphene planes causes strong repulsive interactions between the graphene planes. This then results in coherent ablation of intact atomic planes of the anisotropic crystal.}
   \end{figure}
   \begin{figure}
   \centerline{\includegraphics[width=.3\textwidth]{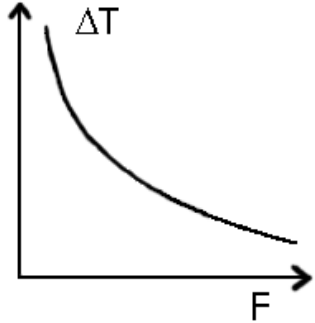}}
   \caption{Schematic illustration of the decrease of critical supercooling $\Delta T = (T_0-T)$ as a function of the laser fluence F (or the fraction of excited electrons). Condensation occurs at temperature T and $T_0(p)$ refers to the thermodynamical phase boundary at pressure p.}
   \end{figure}

   \subsection{Bond Strenghtening}

   Results by Hensel {\it et al.} \cite{6} for Hg may be typical for photoinduced condensation of vapor due to bond strenghtening. The effects  should enhance in nanostructures of Hg when the gap between 6s and 6p states gets smaller, see nonmetal$\rightarrow$transition in Hg clusters \cite{5}. Photoinduced crystallization should also occur close to the liquid/crystal phase boundary.

   Another important case referring to bond strenghtening is the condensation of water, condensation of supercooled or supersaturated vapor, due to photoinduced dipole enhancement \cite{7}. This increases the interaction between the water molecules. Then, for example, an increase of the dipole moment p by a factor 2 might cause a tremendous increase of the rate of condensation, possibly by more than a factor of 10 or even more. Note, approximately (at higher temperature)
   \begin{equation}
                \Gamma \sim \exp ((\Delta G)_k)_{dipole}/kT \sim \exp [a(p)(p^2 p_0^2)/(kT)^2 +..] +... .
   \end{equation}

   Close to the triplepoint, see Fig.3, supercooled or supersaturated vapor may exhibit a transition to ice rather than liquid water, for energetic reasons related to the latent heat of the water/ ice transition.

   Characteristical for the photoinduced
   condensation of water should be its dependence on the laser fluence F. Note, $p=p_0+\Delta p$ and $\Delta p\propto F$  and furthermore
   \begin{equation}
              \ln\Gamma\propto F +...  .
   \end{equation}
   \begin{figure}
   \centerline{\includegraphics[width=.5\textwidth]{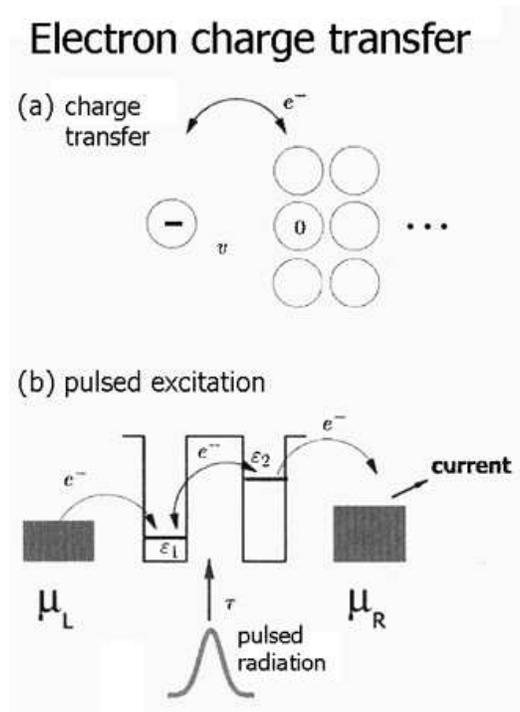}}
   \caption{ Photoassisted (a) electron transfer between chemisorbed atoms, molecules and surface of a solid and (b) electron transport between quantum dots is sketched. The latter may also apply to Cooper--pair tunneling. Of particular interest is the limit where the Cooper pair tunneling, hopping, accompanying charge fluctuations matter. Also of interest is the device for producing ultrafast fluctuating tunnel currents.}
   \end{figure}
   In Fig.9 we illustrate how the occurrence of water condensation (or ice) depends on the change in binding ($\Delta \mu$). Note, as discussed supercooling ($T_0 - T$) decreases for increasing fluence F. Hence, the range of supercooling decreases as q or F increases.

  If also ions are produced upon laser irradiation these play a role and should be included in $(\Delta G)_k$ and $\Gamma$ as discussed before.

   Of course, many thermodynamical processes are affected by photo-induced electron excitations. Not only for graphite, but also for other layered structures one may induce via electron excitations ablation of whole intact atomic planes at the surface, as is the case with graphene.

   Also, for example, one may observe photo--induced demixing of alloys, segregation, continuous and discontinuous changes of bond character (involving a critical fluence) and effects on magnetism. For example, due to electron excitations and its accompanying increase in electron temperature the Curie temperature $T_c$ changes and also depending on the band structure the Heisenberg exchange integral J(T$_{el}$,...) may change from ferromagnetic to antiferromagnetic coupling. In general interesting effects may occur due to photoinduced changes of angular momentum. Then regarding dynamics this is also controlled by angular momentum conservation.

   Note, photo-induced charge transfer, for example between chemi\-sorbed atoms or molecules and surface of solids, and photo-assisted
   tunneling between quantum dots, for illustration see Fig. 10, may yield many interesting results, see Garcia \cite{3}. The interplay of current density and light field, the dependence on light pulse shape and pulse duration should yield interesting results, see Garcia {\it et al.} \cite{3}.

\section{Acknowledgement}
  I thank C. Bennemann for help in preparing this article. Many my interest reactivating useful and enlightening discussions with Profs. H. Baumg\"{a}rtel, M. Garcia and L. W\"{o}ste helped me. I am particularly grateful to P. Stampfli, M. Garcia, H. Jeschke for results.


\begin{thebibliography}{99}

  \bibitem{1} K.H.Bennemann, Ann.Phys. {\bf2}, 475 (Lpz. 2009).

  \bibitem{2} P.Stampfli and K.H.Bennemann, Phys.Rev.B {\bf42}, 7163 (1990), Phys. Rev.B {\bf46}, 10686 (1992), Phys.Rev.B {\bf49}, 7299 (1999).

  \bibitem{3} M.E. Garcia, Habilitation-Thesis , Physik FU-Berlin (1999).

  \bibitem{4} H.Jeschke, Dissertation, Physik FU-Berlin (2000).

  \bibitem{5} G.Pastor, P.Stampfli, and K.H.Bennemann, Europhys. Lett. {\bf7}, 419 (1988); Physica Scripta {\bf38}, 623 (1988).

  \bibitem{6} H.Uchtmann, R.Dettmer, S.D.Baranowskii, and F.Hensel, J.of Chemical Physics {\bf108}, Nr 23, 9775 (1998).

  \bibitem{7} P.Rohwetter et al., Nature Photonics (Advance Online Publ., May, 2010), and L.W\"{o}ste et al., to be publ.(2010).

  \bibitem{8} L.D.Landau, E.M.Lifshitz, Statistical Physics (2. Edition, Pergamon Press); R.Kubo, H.Ichimura, T.Usui, N.Hashitsume,  Thermodynamics
              (Elsevier Publ. Co., N.Y. 1966).

  \bibitem{9} G.S.Zhdanov, Crystal Physics (Academic Press, New York and London, 1965).

  \bibitem{10} L.Pauling, The Nature of the Chemical Bond (3. Edition, Cornell University Press, 1960).

  \bibitem{11} A simple analysis gives approximately $\Delta G_k \simeq (4\pi/3)r_k^2 \sigma +...$, with $r_k = \frac{2\sigma}{\rho[q\Delta T/T + \Delta \mu{'}]}$. Note, for water condensation upon irradiation besides ionization essentially the increase of the dipole coupling might matter
      and then $\Delta \mu{'}$ results from this.



\end{thebibliography}
\end{document}